\documentclass[aps,prb,twocolumn,amsmath,showpacs]{revtex4}
\usepackage{latexsym}
\usepackage{amssymb}
\usepackage{graphicx}
\usepackage{bm}

\newcommand{\pdag}{{\phantom{\dagger}}}
\newcommand{\bq}{\begin{equation}}
\newcommand{\eq}{\end{equation}}
\newcommand{\bn}{\begin{eqnarray}}
\newcommand{\en}{\end{eqnarray}}

\begin{document}

\title{Inelastic cotunneling induced decoherence and relaxation, charge and 
spin currents in an interacting quantum dot under a magnetic field}

\author{Bing Dong}
\affiliation{Department of Physics, Shanghai Jiaotong University,
1954 Huashan Road, Shanghai 200030, China}
\affiliation{Department of Physics and Engineering Physics, Stevens Institute 
of Technology, Hoboken, New Jersey 07030, USA}

\author{Norman J. M. Horing}
\affiliation{Department of Physics and Engineering Physics, Stevens Institute 
of Technology, Hoboken, New Jersey 07030, USA}

\author{H. L. Cui}
\affiliation{Department of Physics and Engineering Physics, Stevens Institute 
of Technology, Hoboken, New Jersey 07030, USA}
\affiliation{School of Optoelectronics Information Science and Technology, 
Yantai University, Yantai, Shandong, China}


\begin{abstract}

We present a theoretical analysis of several aspects of nonequilibirum 
cotunneling through a strong Coulomb-blockaded quantum dot (QD) subject to a 
finite magnetic field in the weak coupling limit. We carry this out by 
developing a generic quantum Heisenberg-Langevin equation approach leading to 
a set of Bloch dynamical equations which describe the nonequilibrium 
cotunneling in a convenient and compact way. These equations describe the time 
evolution of the spin variables of the QD explicitly in terms of the response 
and correlation functions of the free reservoir variables. This scheme not 
only provides analytical expressions for the relaxation and decoherence of the 
localized spin induced by cotunneling, but it also facilitates evaluations of 
the nonequilibrium magnetization, the charge current, and the spin current at 
arbitrary bias-voltage, magnetic field, and temperature. We find that all 
cotunneling events produce decoherence, but relaxation stems only from {\em 
inelastic} spin-flip cotunneling processes. Moreover, our specific 
calculations show that cotunneling processes involving electron transfer (both 
spin-flip and non-spin-flip) contribute to charge current, while spin-flip 
cotunneling processes are required to produce a net spin current in the 
asymmetric coupling case. We also point out that under the influence of a 
nonzero magnetic field, spin-flip cotunneling is an energy-consuming process 
requiring a sufficiently strong external bias-voltage for activation, 
explaining the behavior of differential conductance at low temperature: in 
particular, the splitting of the zero-bias anomaly in the charge current and a 
broad zero-magnitude ``window" of differential conductance for the spin 
current near zero-bias-voltage.

\end{abstract}

\pacs{73.63.Kv, 72.10.Fk, 72.15.Qm, 03.65.Yz}

\today

\maketitle

\section{Introduction}\label{s:I}

Recent advances in probing and manipulating electronic spin in semiconductor 
quantum dots and other nanostructures hold the promise of new applications 
relating to quantum computation and quantum information processing. A single 
electronic spin not only can be used as an elementary quantum memory unit, 
i.e., the qubit, due to its relatively long relaxation time in semiconductors, 
but it is also expected to be useful as an element of calculation in the 
context of quantum computing algorithms or quantum information transport 
processing, which depend essentially on its temporal persistence of quantum 
interference.\cite{Awschalom}

This expectation provides strong motivation to develop a full understanding of 
the coherent evolution dynamics of a single spin in semiconductors. Actually, 
much effort has been made on this matter from both theoretical and 
experimental points of view. In particular, for a single quantum dot (QD), it 
has been predicted that measurements of the tunneling current between two 
leads via this QD may be an appropriate experimental tool to extract 
information about the orientation and dynamics of a single spin localized 
inside the QD.\cite{Bulaevskii2} Indeed, recent scanning tunneling experiments 
have observed modulation of the tunneling current through a single molecule 
with a spin subject to a constant magnetic field at the Larmor frequency, 
which is the characteristic dynamical (precession) frequency of a single spin 
under influence of magnetic field, with a corresponding peak in its noise 
power spectrum.\cite{Manassen,Durkan} This feature has been examined 
theoretically on the basis of two different weak tunneling models in the 
strong Coulomb blockade regime: (a) sequential tunneling using a simple 
quantum rate equation approach,\cite{Mozyrsky} and (b) the two-channel Kondo 
Hamiltonian using the nonequilibrium Green's function formalism jointly with 
the Majorana-fermion representation.\cite{Bulaevskii}

However, from a fundamental quantum mechanical point of view, any quantum 
measurement will inevitably introduce some {\em disturbances} into the 
measured system and consequently induce decoherence in the system variable 
conjugate to the one being measured. Therefore, the information concerning 
spin dynamics extracted from a tunneling measurement is expected to involve a 
reaction signature of the tunneling upon the coherent evolution of the single 
spin. It is therefore crucial to theoretically account for the 
tunneling-measurement-induced spin relaxation and decoherence behaviors as 
functions of temperature and bias-voltage applied between the two leads; so as 
to provide a better understanding of the information obtained from 
measurement, and to give useful insight on how to raise measurement 
efficiency.

The earlier papers cited above have concentrated on interpretation of the peak 
in the current noise spectrum, but a systematic investigation of the 
nonequilibrium relaxation and decoherence effects, as far as we know, is still 
lacking. It is the main purpose of this paper to perform these investigations. 
In treating the open quantum system at hand, we will employ the quantum 
Heisenberg-Langevin equation approach, to establish a set of quantum Bloch 
equations (i.e., equations of motion for the reduced density matrix) for a 
two-level system (a single spin in the QD) tunnel-coupled to two normal leads 
in a fully microscopic way, and then proceed to study the dynamics of a single 
spin qubit in an ambient magnetic field under nonequilibrium transport 
conditions.

As mentioned above, two different tunneling mechanisms have been utilized to 
describe the quantum measurement process. If the chemical potentials of the 
two leads are nearly matched with the energy level of the sandwiched spin, the 
resonance condition is satisfied and the lowest-order tunneling, i.e., 
sequential tunneling, is observed in the transport process. However, it is 
quite likely that the chemical potentials are more probably far from the 
resonant point in actual experiments. In this case, the lowest-order tunneling 
of an electron into the QD is largely suppressed, but at very low temperature, 
a higher-order tunneling mechanism known as {\em inelastic cotunneling} 
dominates the transport in the strong Coulomb blockade regime; in this 
mechanism an electron tunnels from the left lead to a virtual state in the 
dot, and then another electron tunnels from the dot to the right lead without 
changing the charge inside the QD. This is the tunneling mechanism that gives 
rise to the Kondo effect in QD tunneling. In fact, an exact mapping has been 
established between such cotunneling and the anisotropic Kondo problem by 
analyzing and comparing their respective perturbation series for tunneling 
amplitudes.\cite{Matveev} Accordingly, we will adopt the Kondo Hamiltonian in 
this paper to describe the inelastic cotunneling process and study its 
dissipation involved in coherent tunneling via the QD: we do so by developing 
a generic Langevin equation approach in second-order perturbation theory with 
respect to the s-d exchange coupling constant, $J$, in the weak tunneling 
limit.

Moreover, we will evaluate the nonequilibrium spin magnetization of a QD 
subject to a magnetic field in steady state and examine the behavior of charge 
flow (cotunneling current) through the QD within the same framework. Actually, 
an interesting calculation of the spin magnetization of a Kondo QD has already 
been carried out recently by means of the Majorana-fermion Green's function 
technique.\cite{Parcollet} The striking result obtained in that paper is that 
their magnetization result differs from the thermal equilibrium formula even 
at zero order in the spin-leads exchange coupling, $J$. More impressively, 
theoretical analysis shows that if proper account of this nonequilibrium 
magnetization is taken in a calculation of the current, the resulting 
differential conductance will demonstrate double peaks at bias voltages 
$eV=\pm g\mu_{B} B$ (the Zeeman energy), a signature of Kondo effect with a 
constant magnetic field $B$, even in the second-order perturbation 
calculations.\cite{Paaske2} Of course, the log-signature peculiarity of the 
Kondo effect occurs essentially in the next orders of perturbation theory. 
Here, in the present paper, we ignore such higher-order terms; thus we confine 
our study to the ordinary cotunneling processes. A detailed analysis of the 
third-order perturbation contribution to the current has been established in 
Ref.~\onlinecite{Paaske2}, computing an explicit logarithmic enhancement in 
current.

Recent theoretical studies have shown that nonequilibrium Kondo physics is 
fundamentally governed by weak-tunneling perturbation theory when the bias 
voltage is much larger than the Kondo temperature, $T_{\rm K}$; this can be 
ascribed to current induced decoherence of the resonant spin-flip term in 
cotunneling processes, which can eliminate the generic logarithmic divergence 
in conventional Kondo physics.\cite{Kaminski,Rosch,Paaske1} This is the reason 
that the third-order contribution in Ref.~\onlinecite{Paaske2} provides a 
quantitatively relatively small modification to the second-order term in the 
nonequilibrium current formula in the weak tunneling limit (albeit 
qualitatively important). It is also shown in Ref.~\onlinecite{Paaske2} that 
the second-order calculation (cotunneling) of the differential conductance 
exhibits cusps at bias voltages $eV=\pm g\mu_{B} B$, a remnant of the Kondo 
effect. Therefore, we call this cotunneling behavior ``Kondo-type" 
cotunneling. In this paper, we will systematically study this kind of 
nonequilibrium cotunneling through a single spin in a finite magnetic field 
using second-order perturbation theory and will specifically analyze the 
cotunneling processes responsible for its special transport characteristics.

In addition, we will also examine the behavior of the spin current and show 
that {\em inelastic} spin-flip cotunneling can produce a nonzero spin current 
for asymmetric coupling systems subject to a finite magnetic field and an 
activation bias-voltage. Unlike charge current, we find that the sign of spin 
current is independent of the direction of the applied bias-voltage, but it 
does depend on the asymmetry of the coupling constants to the left and right 
leads and the direction of external magnetic field.

The remaining parts of the paper are organized as follows. In 
Sec.~\ref{s:model}, we present the physical model used in this paper: a single 
spin weakly tunnel-coupled to two normal leads. To focus on tunneling induced 
decoherence here, it is assumed that the single spin is free of any other 
dissipative heat bath except for the tunneling reservoirs. In 
Sec.~\ref{s:Lan}, we will then present the derivation of the quantum Langevin 
equations of motion for the single spin. In Sec.~\ref{s:decoherence}, 
qualitative discussion and concrete calculations will be given for the 
resulting decoherence and relaxation rates as functions of magnetic field, 
bias-voltage, and temperature. Sec.~\ref{s:current} focuses first on the 
derivation of closed-form expressions for charge current and spin current 
within the framework of the quantum Langevin equation approach developed here, 
and then addresses all possible cotunneling processes that occur in this 
system, and their respective contributions to the currents. A numerical 
evaluation of differential conductance for the charge current and the spin 
current is provided in the last part of this section. Finally, our conclusions 
are summarized in Sec.~\ref{s:sum}.

\section{Model Hamiltonian}\label{s:model}

We employ the two-lead Kondo Hamiltonian discussed above to model inelastic 
cotunneling through a single spin (or QD) in a magnetic field, $B$, in the 
weak-coupling regime:
\begin{align}
\label{ham}
H= & \, H_0+H_{\rm I}, \ \ \ \ H_{\rm I}=H_{\rm refl}+H_{\rm trans}, \\
H_0= & \, \sum_{\eta {\bf k}\sigma}  \bigr (\varepsilon_{{\bf k}}-\mu_{\eta} 
\bigl ) c^\dag_{\eta
{\bf k} \sigma} c_{\eta {\bf k} \sigma}^\pdag - g\mu_{B} B S^{z}, \cr
H_{\rm refl}  = &\, \sum_{\eta, {\bf k},{\bf k}'} J_{\eta \eta} \bigl [ \bigl( 
c_{\eta {\bf k} \uparrow}^\dag c_{\eta {\bf k}' \uparrow}^\pdag - c_{\eta {\bf 
k} \downarrow}^\dag c_{\eta {\bf k}' \downarrow}^\pdag \bigr) S^z \cr
&\, + c_{\eta {\bf k} \uparrow}^\dag c_{\eta {\bf k}' \downarrow}^\pdag S^{-} 
+ c_{\eta {\bf k} \downarrow}^\dag c_{\eta {\bf k}' \uparrow}^\pdag S^{+} 
\bigr], \cr
H_{\rm trans}  = & \, J_{RL} \sum_{{\bf k},{\bf k}'} \bigl [ \bigl( c_{R {\bf 
k} \uparrow}^\dag c_{L {\bf k}' \uparrow}^\pdag - c_{R {\bf k} 
\downarrow}^\dag c_{L {\bf k}' \downarrow}^\pdag \bigr) S^z \cr
&\, + c_{R {\bf k} \uparrow}^\dag c_{L {\bf k}' \downarrow}^\pdag S^{-} + c_{R 
{\bf k} \downarrow}^\dag c_{L {\bf k}' \uparrow}^\pdag S^{+} \bigr]
 \,\, + \,\, (R \leftrightarrow L), \nonumber
\end{align}
where $c_{\eta {\bf k} \sigma}^\dagger$ ($c_{\eta {\bf k} \sigma}$) creates 
(annihilates) an electron in lead
$\eta$ ($=L,R$) with momentum ${\bf k}$, spin $\sigma$ and bare energy 
$\varepsilon_{{\bf k}}$. $J_{LL}$, $J_{RR}$, and 
$J_{LR}=J_{RL}=\sqrt{J_{LL}J_{RR}}$
are Kondo exchange coupling constants between the electrons and the localized 
spin-$1/2$, ${\bf S}=(S^x,S^y,S^z)$, $S^{\pm}=S^x \pm iS^y$.
$H_0$ stands for the free Hamiltonian containing: (1) two noninteracting 
normal leads, individually in local equilibrium with temperature $T$ (not to 
be confused with decay times to be introduced below), respective chemical 
potentials $\mu_{\eta}$, and Fermi distribution functions defined as 
$f_{\eta}(\epsilon)=[1+ e^{(\epsilon-\mu_{\eta})/k_{B}T}]^{-1}$; and (2) 
Zeeman energy of the localized spin subject to magnetic field $B$ ($g$ and 
$\mu_B$ are the Land\'e factor and the Bohr magneton, respectively). It should 
be noted that we ignore the Zeeman effect in the lead electrons.
The interaction part of the total Hamiltonian, $H_{\rm I}$, also includes two 
terms:
$H_{\rm refl}$ describes the reflection processes, in which an electron from a 
given
lead is scattered back into the {\em same\/} lead in both spin-conserving and 
spin-flip configurations; while $H_{\rm trans}$
describes the transmission events, where an electron from one lead cotunnels 
into the {\em other\/} lead, also in both configurations. Except for the 
tunnel coupling, we ignore all other ``environmental" decay interactions of 
the single spin.

Here, we assume the leads to have a flat density of states,
$\rho_\eta$, in the wide-band limit. We take the chemical potentials, 
$\mu_{\eta}$, to vanish in equilibrium and use this choice as the reference of 
energy throughout the paper. In the nonequilibrium case, we assume the bias 
voltage is applied symmetrically, $\mu_{L}=-\mu_{R}=eV/2$. Throughout, we will 
use units with $\hbar=k_{B}=e=1$.

The conceptual structure of our model is predicated on the idea that the full 
system can be separated into two subsystems: one of which is the measured 
subsystem, the single spin, and the other consists of the two leads jointly 
comprising a ``heat bath" or ``reservoir". The interaction between the two 
subsystems, $H_{\rm I}$, must be weak in order that the separation of the two 
subsystems is physically meaningful. Accordingly, $H_{\rm I}$ generates 
dissipation for the dynamics of the ``open" measured quantum subsystem, which 
is the principal focus for study in this paper. For notational brevity, we 
rewrite this term as a sum of three products of two variables:
\begin{subequations}
\bq
H_{\rm I}=Q^{z}F_{Q^z}+ Q^{+}F_{Q^{+}} + Q^{-}F_{Q^{-}}, \label{Hi}
\eq
with
\begin{align}
Q^{z}=&\, \sum_{\eta,\eta'} Q_{\eta\eta'}^{z}= \sum_{\eta,\eta', {\bf k},{\bf 
k}'} J_{\eta\eta'} \bigl( c_{\eta {\bf k} \uparrow}^\dag c_{\eta' {\bf k}' 
\uparrow}^\pdag - c_{\eta {\bf k} \downarrow}^\dag c_{\eta' {\bf k}' 
\downarrow}^\pdag \bigr), \label{Qz} \\
Q^{+}=&\, \sum_{\eta,\eta'} Q_{\eta\eta'}^{+}= \sum_{\eta, \eta', {\bf k},{\bf 
k}'} J_{\eta\eta'} c_{\eta {\bf k} \uparrow}^\dag c_{\eta' {\bf k}' 
\downarrow}^\pdag, \label{Q+}\\
Q^{-}=&\, \sum_{\eta,\eta'} Q_{\eta\eta'}^{-}= \sum_{\eta, \eta', {\bf k},{\bf 
k}'} J_{\eta\eta'} c_{\eta {\bf k} \downarrow}^\dag c_{\eta' {\bf k}' 
\uparrow}^\pdag, \label{Q-}
\end{align}
as functions of reservoir variables, and the corresponding generalized forces, 
$F_{Q}$, depend on the variables of the measured subsystem as
\bq
F_{Q^z}=S^z,\,\, F_{Q^+}=S^-,\,\, F_{Q^-}=S^+.
\eq
\end{subequations}
Here, the terms $Q^{\pm} F_{Q^{\pm}}$ describe spin-flip tunneling processes, 
in which the conduction electron spin changes its orientation in the process 
of tunneling, and the localized spin is also flipped. On the other hand, the 
term $Q^z F_{Q^z}$ is responsible for non-spin-flip (spin-conserving) 
tunneling, in which no spin exchange takes place. All of these tunneling 
processes are schematically elaborated in Fig.~3.

\section{Quantum Langevin Equations}\label{s:Lan}

In this section, we derive a generic quantum Langevin equation approach and 
establish a set of quantum Bloch equations to describe the dynamics of a 
single spin modeled by Eq.~(\ref{ham}). It is well known that the underlying 
quantum Langevin equation approach has been extensively developed and 
successfully employed in the contexts of quantum electrodynamics and quantum 
optics.\cite{Ackerhalt,Cohen,Milonni} Albeit that the great advantage of this 
scheme is that it allows us to naturally incorporate the effects of quantum 
noise introduced by the ``environment" on the studied system 
variables,\cite{Gardiner} we will take no account of such fluctuation issues 
here. Considering that such noise has a very short correlation time 
(determined by the reservoir correlation time, $\tau_r$), it is reasonable to 
neglect it for the longer time scale ($>\tau_r$) of interest in the present 
paper.

The Heisenberg equations of motion for the spin Pauli operators $S^{z}$, 
$S^{\pm}$ and the lead operators are given by
\begin{align}
i\dot S^z=&\, [S^z, H]_{-}=\, :[S^z,H_{\rm I}]_{-} :\, =\,: \bigl (Q^{-} 
S^{+}- Q^{+} S^{-} \bigr ):\,, \label{eom:sz} \\
i\dot S^{\pm}=&\, [S^{\pm},H]_{-}= \pm \Delta S^{\pm} + :[S^{\pm}, H_{\rm 
I}]_{-}: \cr
=&\,\pm \Delta S^{\pm} \pm : \bigl (2Q^{\pm} S^{z} - Q^z S^{\pm} \bigr ):\,, 
\label{eom:s+} \\
i\dot c_{\eta {\bf k} \uparrow}=&\, [c_{\eta {\bf k} \uparrow}, 
H]_{-}=\varepsilon_{\eta {\bf k}} c_{\eta {\bf k} \uparrow}+ :[c_{\eta {\bf k} 
\uparrow}, H_{\rm I}]_{-}: \cr
=&\, \varepsilon_{\eta {\bf k}} c_{\eta {\bf k} \uparrow}+ \sum_{{\bf k}'} 
\bigl [ S^z \bigl( J_{\eta\eta} c_{\eta {\bf k}' \uparrow} + J_{\eta \bar\eta} 
c_{\bar\eta {\bf k}' \uparrow}\bigr ) \cr
&\,+ S^{-} \bigl ( J_{\eta\eta} c_{\eta {\bf k}' \downarrow}+ J_{\eta\bar\eta} 
c_{\bar\eta {\bf k}' \downarrow} \bigr )\bigr ], \label{eom:cu} \\
i\dot c_{\eta {\bf k} \downarrow}=&\, [c_{\eta {\bf k} \downarrow}, H]_{-} = 
\varepsilon_{\eta {\bf k}} c_{\eta {\bf k} \downarrow}+ :[c_{\eta {\bf k} 
\downarrow}, H_{\rm I}]_{-} : \cr
=&\, \varepsilon_{\eta {\bf k}} c_{\eta {\bf k} \downarrow}+ \sum_{{\bf k}'} 
\bigl [ -S^z \bigl( J_{\eta\eta} c_{\eta {\bf k}' \downarrow} + 
J_{\eta\bar\eta} c_{\bar\eta {\bf k}' \downarrow}\bigr ) \cr
&\,+ S^{-} \bigl ( J_{\eta\eta} c_{\eta {\bf k}' \uparrow}+ J_{\eta\bar\eta} 
c_{\bar\eta {\bf k}' \uparrow} \bigr )\bigr ], \label{eom:cd}
\end{align}
where we have $\varepsilon_{\eta {\bf k}}=\varepsilon_{{\bf k}}-\mu_{\eta}$, 
$\bar\eta=L(R)$ if $\eta=R(L)$, and $\Delta= g\mu_{B}B$. $[A,B]_{\pm}\equiv 
AB\pm BA$ are, respectively, the commutator and the anticommutator of 
operators $A$ and $B$. The equations of motion for $c_{\eta {\bf k} 
\sigma}^\dagger$ are easily obtained by Hermitian conjugation of the equations 
for $c_{\eta {\bf k} \sigma}$. The colon-pair notation, $: (\cdots) :$, in 
these equations denotes {\em normal ordering} of the operators, $\cdots$, 
inside the square brackets: all annihilation reservoir operators $c_{\eta {\bf 
k} \sigma}$ are placed to the right of all spin operators, $S^{z(\pm)}$, and 
the creation reservoir operators $c_{\eta {\bf k} \sigma}^\dagger$ are placed 
to the left of all spin operators, if the operators involved have equal-time 
arguments. For instance, the last two lines in Eqs.~(\ref{eom:cu}) and 
(\ref{eom:cd}) are already normal-ordered. This normal ordering employed here 
is an operator counterpart of determining a cumulant in terms of Feynman 
diagrams with the elimination of disconnected diagrams involving products of 
lower order Green's functions. The latter disconnected diagram terms involve 
the effects of weak coupling (to the bath) which oscillate rapidly at the high 
frequencies of microscopic dynamics, with attendant destructive interference. 
While such terms do contribute small quantum corrections (``renormalization", 
``radiative corrections") to the microscopic dynamics on a short time scale, 
they are negligible in the context of the much longer time scale implicitly 
under consideration in our formulation of a quantum Heisenberg-Langevin 
equation. A full explanation of the normal ordering scheme in the equations of 
motion is provided in Refs.~\onlinecite{Ackerhalt}--\onlinecite{Haake}, to 
which we refer the reader.

Formally integrating these Heisenberg equations of motion from initial time 
$0$ to $t$ we obtain the exact solutions for these operators as
\begin{subequations}
\label{solution}
\begin{align}
S^{z}(t)=&\, S^{z}(0)-i\int_{0}^t dt' : [S^z(t'), H_{\rm I}(t')]_{-} :\,, \\
S^{\pm}(t)=&\, e^{\mp i \Delta t} S^{\pm}(0) -i \int_{0}^t dt' e^{\mp i \Delta 
(t-t')} \cr
&\, \times : [S^{\pm}(t'), H_{\rm I}(t')]_{-} :\,, \\
c_{\eta {\bf k} \sigma}(t)=&\,e^{-i\varepsilon_{\eta {\bf k}}t} c_{\eta {\bf 
k} \sigma}(0) -i \int_{0}^t dt' e^{-i \varepsilon_{\eta {\bf k}}(t-t')} \cr
&\, \times :[c_{\eta {\bf k} \sigma}(t'), H_{\rm I}(t')]_{-} :\,.
\end{align}
\end{subequations}
In the absence of interaction, $H_{\rm I}\rightarrow 0$, these solutions 
become:
\begin{subequations}
\label{free}
\begin{align}
S_o^z(t)=&\,S_o^z(t'), \label{free:sz} \\
S_o^{\pm}(t)=&\, e^{\mp i \Delta (t-t')} S_o^{\pm}(t'), \label{free:s+} \\
c_{\eta {\bf k} \sigma}^o(t)=&\, e^{-i\varepsilon_{\eta {\bf k}}(t-t')} 
c_{\eta {\bf k} \sigma}^o(t'). \label{free:c}
\end{align}
\end{subequations}
A standard assumption in the derivation of a quantum Langevin equation is that 
the time scale of decay processes is much slower than that of free evolution, 
which is reasonable in the weak-tunneling approximation. In this context it is 
appropriate to substitute the time-dependent decoupled reservoir and spin 
operators of Eq.~(\ref{free}) into the formal solutions of 
Eq.~(\ref{solution}). Obviously, the full solution for the reservoir operator 
comprises two contributions, one from free evolution and the other from 
reaction of the spin through the weak coupling, and we denote these with 
superscripts ``$o$" and ``$i$", respectively:
\begin{subequations}
\label{coi}
\bq
c_{\eta {\bf k} \sigma}(t)=c_{\eta {\bf k} \sigma}^o(t)+ c_{\eta {\bf k} 
\sigma}^i(t), \label{coi1}
\eq
with
\bq
c_{\eta {\bf k} \sigma}^i(t)=-i \int_{0}^t dt' :[c_{\eta {\bf k} \sigma}^o(t), 
H_{\rm I}^o(t')]_{-}:\,,
\eq
\end{subequations}
where $H_{\rm I}^o$ is composed of the operators in $H_{\rm I}$ which are 
replaced by their decoupled counterparts (interaction picture). In fact, this 
is just the operator formulation of linear response theory. It should also be 
noted that Eq.~(\ref{coi1}) implies that the two subsystems, the quantum dot 
and the reservoirs, are completely isolated before $t_0=0$, and the 
perturbative interaction, $H_{\rm I}$, is adiabatically switched on from the 
initial time $t=t_0$. Using Eq.~(\ref{coi}), the reservoir variables, 
$Q_{\eta\eta'}^{z(\pm)}(t)$, become (Appendix \ref{app:a}):
\begin{subequations}
\label{Q}
\begin{align}
Q_{\eta\eta'}^z(t)=&\, Q_{\eta\eta'}^{zo}(t) + Q_{\eta\eta'}^{zi}(t) \cr
=&\, Q_{\eta\eta'}^{zo}(t) -i\theta(\tau)\int_{-\infty}^t d\tau :[Q_{\eta 
\eta'}^{zo}(t), H_{\rm I}^o(t')]_{-}:, \cr
& \label{Q:qz} \\
Q_{\eta\eta'}^{\pm}(t)=&\, Q_{\eta\eta'}^{\pm o}(t) + Q_{\eta\eta'}^{\pm i}(t) 
\cr
=&\, Q_{\eta\eta'}^{\pm o}(t) -i\theta(\tau)\int_{-\infty}^t d\tau :[Q_{\eta 
\eta'}^{\pm o}(t), H_{\rm I}^o(t')]_{-}:, \label{Q:q+}
\end{align}
\end{subequations}
with $\tau=t-t'$ and $\theta(\tau)$ represents the Heaviside step-function.
Similarly, the formal solutions for the spin operators are also divided into 
two parts:
\begin{subequations}
\label{S}
\bn
S^{z}(t)&=& S_{o}^{z}(t)+S_{i}^{z}(t) \cr
&=& S_{o}^{z}(t)-i\theta(\tau) \int_{-\infty}^t d\tau :[S_{o}^z(t), H_{\rm 
I}^o(t')]_{-}:\,, \cr
&& \\
S^{\pm}(t)&=& S_{o}^{\pm}(t) + S_{i}^{\pm}(t) \cr
&=& S_{o}^{\pm}(t)-i\theta(\tau) \int_{-\infty}^t d\tau :[S_{o}^{\pm}(t), 
H_{\rm I}^o(t')]_{-}:\,. \cr
&&
\en
\end{subequations}

Substituting these approximate solutions of Eqs.~(\ref{Q}) and (\ref{S}) into 
the equations of motion for $S^z$ and $S^{\pm}$ [Eqs.~(\ref{eom:sz}) and 
(\ref{eom:s+})] and taking average evaluations with respect to the reservoir 
electron ensemble $\langle \cdots \rangle_{e}$ and over the localized spin 
degrees of freedom $\langle \cdots \rangle_{s}$, one can derive the desired 
quantum Bloch equations up to second order in the Kondo coupling constant $J$. 
After some algebraic manipulations (details are provided in Appendix 
\ref{app:a}), the quantum dynamic equations take the compact form:
\begin{widetext}
\bn
\langle \dot S^{z}\rangle &=& -{1\over 2} \theta(\tau) \int_{-\infty}^t d\tau 
\langle [Q_{o}^{-}(t), Q_{o}^{+}(t')]_{+} \rangle_{e} \langle [S^{+}(t), 
F_{Q^+}(t')]_{-} \rangle_{s} \cr
&& -{1\over 2} \theta(\tau) \int_{-\infty}^t d\tau \langle [Q_{o}^{-}(t), 
Q_{o}^{+}(t')]_{-} \rangle_{e} \langle [S^{+}(t), F_{Q^+}(t')]_{+} \rangle_{s} 
\cr
&& +{1\over 2} \theta(\tau) \int_{-\infty}^t d\tau \langle [Q_{o}^{+}(t), 
Q_{o}^{-}(t')]_{+} \rangle_{e} \langle [S^{-}(t), F_{Q^-}(t')]_{-} \rangle_{s} 
\cr
&& + {1\over 2} \theta(\tau) \int_{-\infty}^t dt' \langle [Q_{o}^{+}(t), 
Q_{o}^{-}(t')]_{-} \rangle_{e} \langle [S^{-}(t), F_{Q^-}(t')]_{+} 
\rangle_{s}, \label{eom:sz1}
\en
\bn
\langle \dot S^{\pm}\rangle &=& \mp i \Delta \langle S^\pm \rangle \mp 
\theta(\tau) \int_{-\infty}^t d\tau \langle [Q_{o}^{\pm}(t), 
Q_{o}^{\mp}(t')]_{+} \rangle_{e} \langle [S^{z}(t), F_{Q^\mp}(t')]_{-} 
\rangle_{s} \cr
&& \mp \theta(\tau) \int_{-\infty}^t d\tau \langle [Q_{o}^{\pm}(t), 
Q_{o}^{\mp}(t')]_{-} \rangle_{e} \langle [S^{z}(t), F_{Q^\mp}(t')]_{+} 
\rangle_{s} \cr
&& \pm {1\over 2} \theta(\tau) \int_{-\infty}^t d\tau \langle [Q_{o}^{z}(t), 
Q_{o}^{z}(t')]_{+} \rangle_{e} \langle [S^{\pm}(t), F_{Q^z}(t')]_{-} 
\rangle_{s} \cr
&& \pm {1\over 2} \theta(\tau) \int_{-\infty}^t d\tau \langle [Q_{o}^{z}(t), 
Q_{o}^{z}(t')]_{-} \rangle_{e} \langle [S^{\pm}(t), F_{Q^z}(t')]_{+} 
\rangle_{s}. \label{eom:s+1}
\en
\end{widetext}
In these equations, we drop the superscript ``$o$" in the spin operators 
occurring inside integrations, since they involve the dynamical spin variables 
after taking expectation values. However, it must be borne in mind that their 
time evolutions are governed by Eqs.~(\ref{free:sz}) and (\ref{free:s+}). 
Apart from free evolution, it is clear that the spin dynamics are modified by 
the spin-lead interaction in a way that is precisely relevant to the response 
function, $R^{ab}(t,t')$, and correlation function, $C^{ab}(t,t')$, 
($a,b=z,+,-$) of free reservoir variables, which are defined as:
\bn
R^{ab}(t,t')&=& {1\over 2}\theta(\tau) \langle [Q_{o}^{a}(t), 
Q_{o}^{b}(t')]_{-} \rangle_{e}, \label{responsef}\\
C^{ab}(t,t')&=& {1\over 2}\theta(\tau) \langle [Q_{o}^{a}(t), 
Q_{o}^{b}(t')]_{+} \rangle_{e}. \label{correlationf}
\en
These forms of quantum Langevin-type dynamic equations, expressed explicitly 
in terms of the correlation and response functions of free reservoir 
variables, have also been proposed in Ref.~\onlinecite{Smirnov} by employing 
the quantum Furutsu-Novikov theorem. The present derivation seems more direct 
and its meaning is more transparent.

Considering the reservoirs to be in separate (local) equilibrium states except 
for differing chemical potentials (reflecting the bias-voltage driving the 
current) and noting that these free fermion reservoir operators, $c_{\eta {\bf 
k} \sigma}^o$, $c_{\eta {\bf k} \sigma}^{o \dagger}$, obey Wick's theorem 
(without correlation between the leads), we can readily express the functions 
$R^{ab}(t,t')$ and $C^{ab}(t,t')$ in terms of reservoir distribution 
functions. The calculational details are provided in Appendix \ref{app:b}. 
Here we cite some useful properties. Firstly, these response and correlation 
functions are functions only of the time difference $\tau=t-t'$. Secondly, 
these functions are related as
\bn
R(\tau)&=&R^{+-}(\tau)=R^{-+}(\tau)={1\over 2}R^{zz}(\tau), \label{rtau} \\
C(\tau)&=&C^{+-}(\tau)=C^{-+}(\tau)={1\over 2}C^{zz}(\tau). \label{ctau}
\en
Therefore, it is convenient to introduce single Fourier time transforms for 
the two bath functions into frequency space:
\bn
R(\omega)&=& \int_{-\infty}^\infty d\tau e^{i\omega \tau} R(\tau), \\
C(\omega)&=& \int_{-\infty}^\infty d\tau e^{i\omega \tau} C(\tau).
\en
Thirdly, the spectral function $C(\omega)$ is an even function of $\omega$, 
while the imaginary part of the frequency-dependent retarded susceptibility 
$R(\omega)$ is an odd function. In equilibrium, they are exactly related by 
the fluctuation-dissipation theorem.

Employing the definitions of response and correlation functions and free 
evolution relation $S^{\pm}(t')=e^{\pm i\Delta \tau} S^{\pm}(t)$, 
Eq.~(\ref{eom:sz1}) yields
\bn
\dot S^{z}&=& -2 S^{z} \int_{-\infty}^t d\tau e^{-i\Delta \tau} C(\tau) - 
\int_{-\infty}^t d\tau e^{-i\Delta \tau} R(\tau) \cr
&&-2 S^{z} \int_{-\infty}^t d\tau e^{i\Delta \tau} C(\tau) + \int_{-\infty}^t 
d\tau e^{i\Delta \tau} R(\tau).
\en
(Hereafter, we suppress the brackets around the spin variables since they are 
all $c$-numbers.) In a transport measurement experiment, a single spin decays 
to its external bias-voltage-driven steady state in a characteristic time, 
$\tau_c$, of the system. If we assume that the single spin changes 
significantly only over a time scale $t\gg \tau_c$, an appropriate Markov 
approximation may be generated by making the replacement
\bq
\int_{-\infty}^{t} d\tau \Longrightarrow \int_{-\infty}^{\infty} d\tau.
\eq
From a physical point of view, this presumption is consistent with the normal 
ordering scheme performed in the operator equations of motion, in regard to 
elimination of the rapid oscillations in microscopic dynamics. In this case, 
the equation of motion for $S^z$ can be further simplified as
\bn
\dot S^{z}&=& -2 [C(\Delta) + C(-\Delta)]\, S^{z} + R(\Delta) - R(-\Delta) \cr
&=& -4C(\Delta) \, S^{z} + 2R(\Delta). \label{sz}
\en
Analogously, the quantum Langevin equation for $S^{\pm}$ becomes
\bn
\dot S^{\pm}&=& \mp i \Delta\, S^\pm \cr
&& - 2 S^{\pm} \int_{-\infty}^t d\tau e^{\pm i\Delta \tau} C(\tau) - 2 S^{\pm} 
\int_{-\infty}^t d\tau C(\tau) \cr
&=& \mp i \Delta\, S^\pm -2 [C(\Delta)+ C(0)]\, S^{\pm}. \label{s+}
\en

\section{Nonequilibrium magnetization, decoherence and 
relaxation}\label{s:decoherence}

In this section, on the basis of the derived Bloch equations, Eqs.~(\ref{sz}) 
and (\ref{s+}), we will carry out analytical evaluations of the relaxation and 
decoherence rates, as well as the magnetization of the single spin under 
transport conditions, as functions of temperature, bias-voltage, and external 
magnetic field.

It is well known that there are two decay mechanisms leading to standard Bloch 
equations which define two distinct relaxation time scales: (1) The 
longitudinal relaxation time, $T_1$, is responsible for the spin magnetic 
moment relaxation, while (2) the transverse relaxation time, $T_2$, is 
responsible for decoherence of the quantum superposition state composed of the 
two spin states $\sigma=\uparrow$ and $\downarrow$. These time scales are 
defined by the time evolutions of $S^z(t)$ and $S^\pm$, respectively:
\bn
{1 \over T_1} &=& 4C(\Delta)=2\pi \left ( J_{LL}^2 \rho_L^2+ J_{RR}^2 \rho_R^2 
\right ) T \varphi \left ( \frac{\Delta}{T} \right ) \cr
&& \hspace{-0.8cm}+ 2\pi J_{LR}^2 \rho_{L}\rho_{R} T \left [ \varphi \left ( 
\frac{\Delta + V}{T}\right ) + \varphi \left ( \frac{\Delta- V}{T}\right ) 
\right ], \\
{1 \over T_2} &=& 2[C(\Delta) + C(0)] = {1 \over 2T_1}+ 2C(0) \cr
&=& {1 \over 2T_1} + 2\pi \left ( J_{LL}^2 \rho_L^2+ J_{RR}^2 \rho_R^2 \right 
) T \cr
&& + 2\pi J_{LR}^2 \rho_{L}\rho_{R} T \varphi \left ( \frac{V}{T}\right ).
\en
In deriving these results, we employed Eqs.~(\ref{correlation}) and 
(\ref{phi}). It is noteworthy that the transverse spin relaxation rate, 
$1/T_2$, includes two contributions: the relaxation-induced dephasing, 
$1/2T_1$, and also pure decoherence, $2C(0)$.

From Eqs.~(\ref{eom:sz1}) and (\ref{eom:s+1}), we can easily deduce that the 
longitudinal relaxation time ($T_1$) stems completely from spin-flip 
cotunneling events, which is conceptually consistent with the physical 
definition of spin relaxation and implies its dependence on magnetic field. 
Furthermore, spin-flip processes also contribute to decoherence with the 
partial rate, $1/2T_1$. In contrast, non-spin-flip processes do not induce 
spin relaxation but they do contribute to pure decoherence with the partial 
rate $2C(0)$, which is independent of magnetic field. This difference in the 
magnetic field dependences of the two rates may be understood in the following 
terms: the non-spin-flip process entails charge transport through the QD via a 
virtual state but the QD eventually returns back to its original spin state 
without changing energy (which is why this process is referred to as {\em 
elastic} cotunneling in the literature); whereas energy exchange does take 
place between the QD and leads in the spin-flip process in a finite magnetic 
field, in which the spin of the QD is finally flipped and thus the QD is {\em 
inelastically} excited or decays accompanied by excess energy, the Zeeman 
energy, $\Delta$. Of course, in the absence of an external magnetic field, 
spin-flip cotunneling also becomes {\em elastic}. In this case, the two 
relaxation times are equal (i.e., it is inelastic cotunneling that makes them 
differ),
\bn
{1 \over T_1^0}={1 \over T_2^0}&=&4C(0)=4\pi \left ( J_{LL}^2 \rho_L^2+ 
J_{RR}^2 \rho_R^2 \right ) T \cr
&& + 4\pi J_{LR}^2 \rho_{L}\rho_{R} T \varphi \left ( \frac{V}{T}\right ).
\en

On the other hand, in the limit of zero bias-voltage (equilibrium condition), 
the relaxation rates become
\bq
{1 \over T_1^{\rm eq}} = 4\pi \left ( \frac{J_{LL}^2 \rho_L^2+ J_{RR}^2 
\rho_R^2}{2} + J_{LR}^2 \rho_{L}\rho_{R} \right ) T \varphi \left ( 
\frac{\Delta}{T} \right ),
\eq
\bq
{1 \over T_2^{\rm eq}} = {1 \over 2T_1^{\rm eq}} + 4\pi \left ( \frac{J_{LL}^2 
\rho_L^2+ J_{RR}^2 \rho_R^2}{2} + J_{LR}^2 \rho_{L}\rho_{R} \right ) T. 
\label{T2eq}
\eq
It is clear that the contribution of non-spin-flip cotunneling to the pure 
decoherence rate, the second term in Eq.~(\ref{T2eq}), is proportional to 
temperature, while spin-flip cotunneling leads to a somewhat complicated 
temperature dependence, $\Delta\coth(\Delta/2T)$. This difference can also be 
ascribed to energy exchange in the dissipation process. It is worth noting 
that without transport ($V\rightarrow 0$), dissipation (relaxation and 
decoherence) is due solely to quantum thermal fluctuations (thermal noise). 
Furthermore, if the external magnetic field is quenched, the thermal 
fluctuations are purely non-energy-consuming:
\bq
\tau_{\rm th}={1 \over T_1^{eq}}={1 \over T_2^{eq}}=8\pi \left ( 
\frac{J_{LL}^2 \rho_L^2+ J_{RR}^2 \rho_R^2}{2}+ J_{LR}^2 \rho_{L}\rho_{R} 
\right ) T, \label{Tth}
\eq
indicating that the dissipation is totally determined by thermodynamics 
(temperature) of the reservoirs.

For illustrative purposes, we exhibit in Fig.~1 the dependences of the 
relaxation rate and the dephasing rate on magnetic field (a,b) and on 
bias-voltage (c,d) for given temperatures. At relatively low temperatures, 
$T/V=0.01$ (or $T/\Delta=0.01$), these rates show linear increase with respect 
to bias-voltage $V$ (magnetic field $\Delta$) with the rates of increase 
depending on the relative magnitudes of $V$ and $\Delta$. Interestingly, 
dephasing, $1/T_2$, is independent of $V$ for $V< \Delta$, as shown in 
Fig.~1(c). This comes about because the hyperbolic cotangent functions behave 
as $\varphi((\Delta+V)/T)+ \varphi((\Delta-V)/T) \rightarrow |\Delta+V| + 
|\Delta-V|$ in the limit $T\rightarrow 0$. As expected, rising temperature 
smears out the low-temperature structures in these rates [Figs.~1(b) and (c)]. 
Finally, the temperature dependences of the two rates are summarized in 
Fig.~2.

The magnetization of the QD, defined as $M=S^z$, is readily obtained using the 
steady solution of Eq.~(\ref{sz}) as
\begin{widetext}
\bq
M(\Delta,V)=S_{\infty}^z=\frac{R(\Delta)}{2C(\Delta)} =\frac{\displaystyle 
\left ( \frac{J_{LL}^2 \rho_L^2 + J_{RR}^2 \rho_R^2}{2} + J_{LR}^2 \rho_{L} 
\rho_{R} \right ) \frac{\Delta}{T}}{\displaystyle \left ( J_{LL}^2 \rho_L^2+ 
J_{RR}^2 \rho_R^2 \right ) \varphi \left ( \frac{\Delta}{T} \right ) + 
J_{LR}^2 \rho_{L}\rho_{R} \left [ \varphi \left ( \frac{\Delta + V}{T}\right ) 
+ \varphi \left ( \frac{\Delta- V}{T}\right ) \right ]}, \label{magnetization}
\eq
\end{widetext}
which is identical to previous theoretical result.\cite{Parcollet,Paaske2} In 
absence of bias-voltage, $V=0$, it reduces to the equilibrium expression 
$M(\Delta,0)={1\over 2} \tanh(\Delta/2T)$.

\begin{figure}[htb]
\includegraphics [width=8.5cm,height=9cm,angle=0,clip=on]{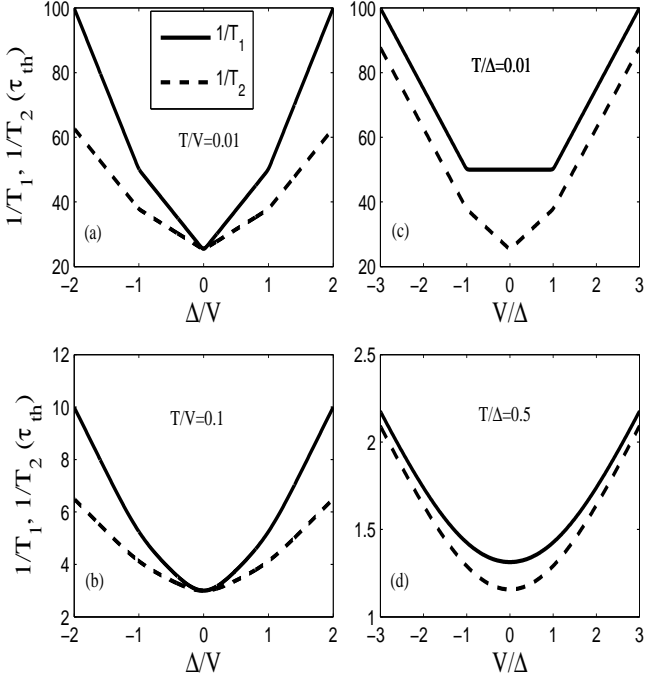}
\caption{The cotunneling-induced spin relaxation rate ($T_1^{-1}$) and 
decoherence rate ($T_2^{-1}$) as functions of magnetic field $\Delta$ (a,b) 
and of bias-voltage $V$ (c,d) for given temperatures indicated in these 
figures. The parameters we use in calculation are: $J_{LL} 
\rho_L=J_{RR}\rho_R=J_{LR}\sqrt{\rho_L \rho_R}= 0.02$. Here, we use the decay 
rate of purely thermal fluctuations, $\tau_{\rm th}$, Eq.~(\ref{Tth}), as the 
unit of the two rates.} \label{fig1}
\end{figure}

\begin{figure}[htb]
\includegraphics [width=7cm,height=7cm,angle=0,clip=on]{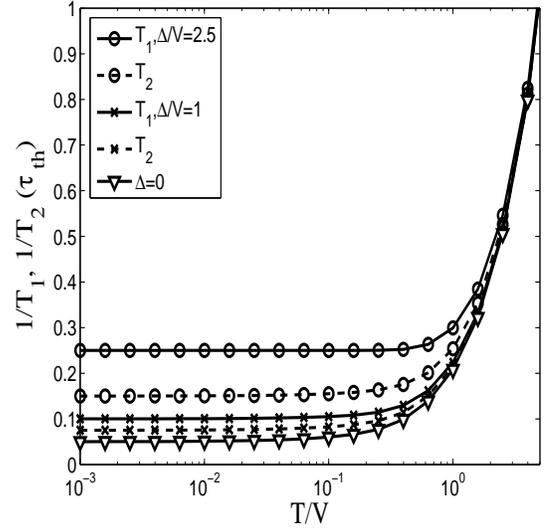}
\caption{The cotunneling-induced spin relaxation rate ($T_1^{-1}$) and 
decoherence rate ($T_2^{-1}$) as functions of temperature for $\Delta/V=2.5$, 
$1.0$, and $0$. Other parameters are the same as in Fig.~1.} \label{fig2}
\end{figure}

\section{Nonlinear Tunneling Current}\label{s:current}

The calculation of steady state tunneling charge current, $I^c$, measuring the 
charge flow from left lead to right lead, is based on the equation of motion 
for the charge density $N_{L}=\sum_{\sigma} N_{L\sigma}= \sum_{{\bf k}} c_{L 
{\bf k} \sigma}^\dagger c_{L {\bf k} \sigma}^\pdag$ in the left lead, which is 
a sum of both spin-up and spin-down electrons flows, $I^c=I_{L\uparrow} + 
I_{L\downarrow}$,
\bn
I_{L\uparrow} &=&  -\langle \dot N_{L\uparrow} \rangle = i  \langle 
[N_{L\uparrow}, H]\rangle \cr
&=& i \langle (Q_{LR}^{z\uparrow\uparrow}- Q_{RL}^{z\uparrow\uparrow}) S^z - 
(Q_{LL}^- + Q_{RL}^- ) S^+ \cr
&& + (Q_{LL}^+ + Q_{LR}^+) S^- \rangle, \label{iup}
\en
\bn
I_{L\downarrow} &=&  -\langle \dot N_{L\downarrow} \rangle = i  \langle 
[N_{L\downarrow}, H]\rangle \cr
&=& i \langle (Q_{RL}^{z\downarrow\downarrow}- Q_{LR}^{z\downarrow\downarrow}) 
S^z + (Q_{LL}^- + Q_{LR}^- ) S^+ \cr
&& - (Q_{LL}^+ + Q_{RL}^+) S^- \rangle, \label{idown}
\en
with the definition $Q_{\eta\eta'}^{z\sigma\sigma}=J_{\eta\eta'}\sum_{{\bf k}, 
{\bf k}'} c_{\eta {\bf k} \sigma}^{\dagger} c_{\eta' {\bf k}' 
\sigma}^{\pdag}$.
Using the same procedure described in the preceding section and Appendix 
\ref{app:a}, and employing the various response and correlation functions of 
the free reservoir variables determined in Appendix \ref{app:b}, we have 
derived analytic expressions for the spin-resolved current, $I_{L\sigma}$. For 
example, the spin-up current, $I_{L\uparrow}$, takes the form
\bn
I_{L\uparrow}&=& {1\over 2} \int_{-\infty}^t d\tau \left 
[R_{LR,RL}^{zz\uparrow\uparrow}(\tau)- R_{RL,LR}^{zz\uparrow\uparrow}(\tau) 
\right ] \cr
&& -  \int_{-\infty}^t d\tau  \left \{ e^{-i \Delta \tau} \left 
[R_{LL,LL}^{-+}(\tau) + R_{RL,LR}^{-+}(\tau) \right ] \right. \cr
&& \left. - e^{i\Delta \tau} \left [ R_{LL,LL}^{+-}(\tau) + R_{LR,RL}^{+-} 
\right ] \right \} \cr
&& - 2  S^z \int_{-\infty}^t d\tau \left \{ e^{-i \Delta \tau} \left [ 
C_{LL,LL}^{-+}(\tau)+ C_{RL,LR}^{-+}(\tau) \right ] \right.\cr
&& \left. + e^{i\Delta \tau} \left [C_{LL,LL}^{+-}(\tau) + 
C_{LR,RL}^{+-}(\tau) \right ] \right \}.
\en
Using Eqs.~(\ref{crzz1}), (\ref{crzz2}), (\ref{cr+-1})--(\ref{cr+-3}), and 
then making the replacement $\int_{-\infty}^t d\tau \Rightarrow 
\int_{-\infty}^{\infty} d\tau$, we finally arrived at an explicit result for 
$I_{L\uparrow}$ as a function of temperature and bias-votage [after performing 
the $\epsilon$-integrals of Eqs.~(\ref{fermi:integral1}) and 
(\ref{fermi:integral2})],
\bn
I_{L\uparrow}&=&  {3 \pi \over 2} J_{LR}^2 \rho_L \rho_R V + \pi \Delta \left 
( J_{LL}^2 \rho_L^2 + J_{LR}^2 \rho_L \rho_R \right )  \cr
&-&  2\pi T \left [ J_{LL}^2 \rho_L^2 \varphi \left ( {\Delta \over T}\right ) 
+ J_{LR}^2 \rho_L \rho_R \varphi \left ( {\Delta+V \over T} \right )  \right ] 
S^z.\cr
&&
\en
Similarly, the current from spin-down electrons, $I_{L\downarrow}$, is given 
by
\bn
I_{L\downarrow}&=& { 3\pi \over 2} J_{LR}^2 \rho_L \rho_R V - \pi \Delta \left 
( J_{LL}^2 \rho_L^2 + J_{LR}^2 \rho_L \rho_R \right )  \cr
&+&  2\pi T \left [ J_{LL}^2 \rho_L^2 \varphi \left ( {\Delta \over T}\right ) 
+ J_{LR}^2 \rho_L \rho_R \varphi \left ( {\Delta-V \over T} \right ) \right ] 
S^z.\cr
&&
\en
The total charge current, $I^c$, is thus
\bn
I^c&=& \pi J_{LR}^2 \rho_L \rho_R \left \{ 3 V + 2 S^z T \left [ \varphi \left 
( {\Delta-V \over T} \right ) \right. \right. \cr
&& \left. \left. - \varphi \left ( {\Delta+V \over T} \right ) \right ] \right 
\}. \label{ic}
\en
It is worth noting that the cotunneling current, Eq.~(\ref{ic}), is just 
proportional to second-order in the exchange coupling constant, $J_{LR}$, and  
higher-order contributions are all neglected. This is because we employ the 
approximation formula, Eq.~(\ref{qasb}), to derive non-Markovian quantum 
dynamic equations and the current, leading to the absence of the 
characteristic logarithmic divergence term in current. Thus, the present 
approach can \emph{not} be applied to describe strong Kondo correlations, but 
can be used to study the ordinary cotunneling process in the weak 
tunnel-coupling limit.

Due to the fact that spin-up electrons are coupled with spin-down electrons 
via the spin-flip processes in this model, there is an imbalance between the 
spin-up current and spin-down current, i.e., there is a net spin current, 
$I^s$, with respect to the left lead, defined as
\bn
I^s&=& I_{L\uparrow}- I_{L\downarrow}=2\pi \Delta ( J_{LL}^2 \rho_{L}^2 + 
J_{LR}^2 \rho_L \rho_R ) \cr
&& - 2 \pi T \left \{ 2 J_{LL}^2 \rho_{L}^2 \varphi \left ( {\Delta \over T} 
\right ) + J_{LR}^2 \rho_L \rho_R \left [ \varphi \left ( {\Delta-V \over T} 
\right ) \right. \right. \cr
&& \left. \left. + \varphi \left ( {\Delta+V \over T} \right ) \right ] \right 
\} S^z. \label{is}
\en
To better understand these formulae and the physical perspective involved, we 
elaborate the physical picture of cotunneling processes through a QD in a 
finite magnetic field. When the electronic levels of the QD, 
$\epsilon_{d\sigma}$, are far below the chemical potentials of the two leads, 
i.e. $\epsilon_{d\sigma} \ll \mu_{L(R)}$, the first-order tunneling process, 
sequential tunneling, vanishes. However, higher-order tunneling processes, 
{\em cotunneling}, are active and dominate the quantum transport.
In the strong Coulomb blockade regime, the QD is always {\em singly} occupied 
by an electron because of the deep electronic energy, $\epsilon_{d\sigma}$, 
and the extremely strong charging energy, $U\rightarrow \infty$, involved when 
an additional, excess electron attempts to enter the QD, i.e. 
$\epsilon_{d\sigma} +U \gg \mu_{L(R)}$. This steady occupation means that {\em 
no charge-fluctuation} takes place, provided that the applied bias-voltage is 
not strong enough to force the chemical potential in one of leads below the QD 
level, $\epsilon_{d\sigma}$, so as to drive the transport into the sequential 
tunneling regime. Therefore, in nonequilibrium conditions, a cotunneling event 
consists of two single-particle tunneling processes, 
\textcircled{\footnotesize 1} and \textcircled{\footnotesize 2}, which can 
take place in sequence as follows: event \textcircled{\footnotesize 1}, an 
electron inside the QD with spin-$\sigma$ will at first tunnel out to a lead L 
or R, inducing a virtual empty state in the dot, but this is immediately 
followed by a second single-particle tunneling event 
\textcircled{\footnotesize 2} in which an electron in one of the leads is 
injected into the QD with the same spin-$\sigma$ (spin-conserving elastic 
cotunneling) or with the opposite spin (spin-flip inelastic cotunneling). {\em 
The two tunneling events occur via a virtual empty-dot-state in a very short 
time interval to insure coherence.} Importantly, spin-flip cotunneling 
provides a mechanism for the spin orientation of the QD to be changed, which 
is fully quantum phenomenology. Moreover, it stimulates intrinsic {\em 
spin-fluctuation}, which is a fundamental concept in the context of Kondo 
physics.
Obviously, there is a total of $16$ different cotunneling events allowed in 
this strong Coulomb blockade system, which are schematically shown in Fig.~3. 
We can classify them in four different categories/types.

\begin{figure*}[htb]
\includegraphics [width=18cm,height=17.5cm,angle=0,clip=on]{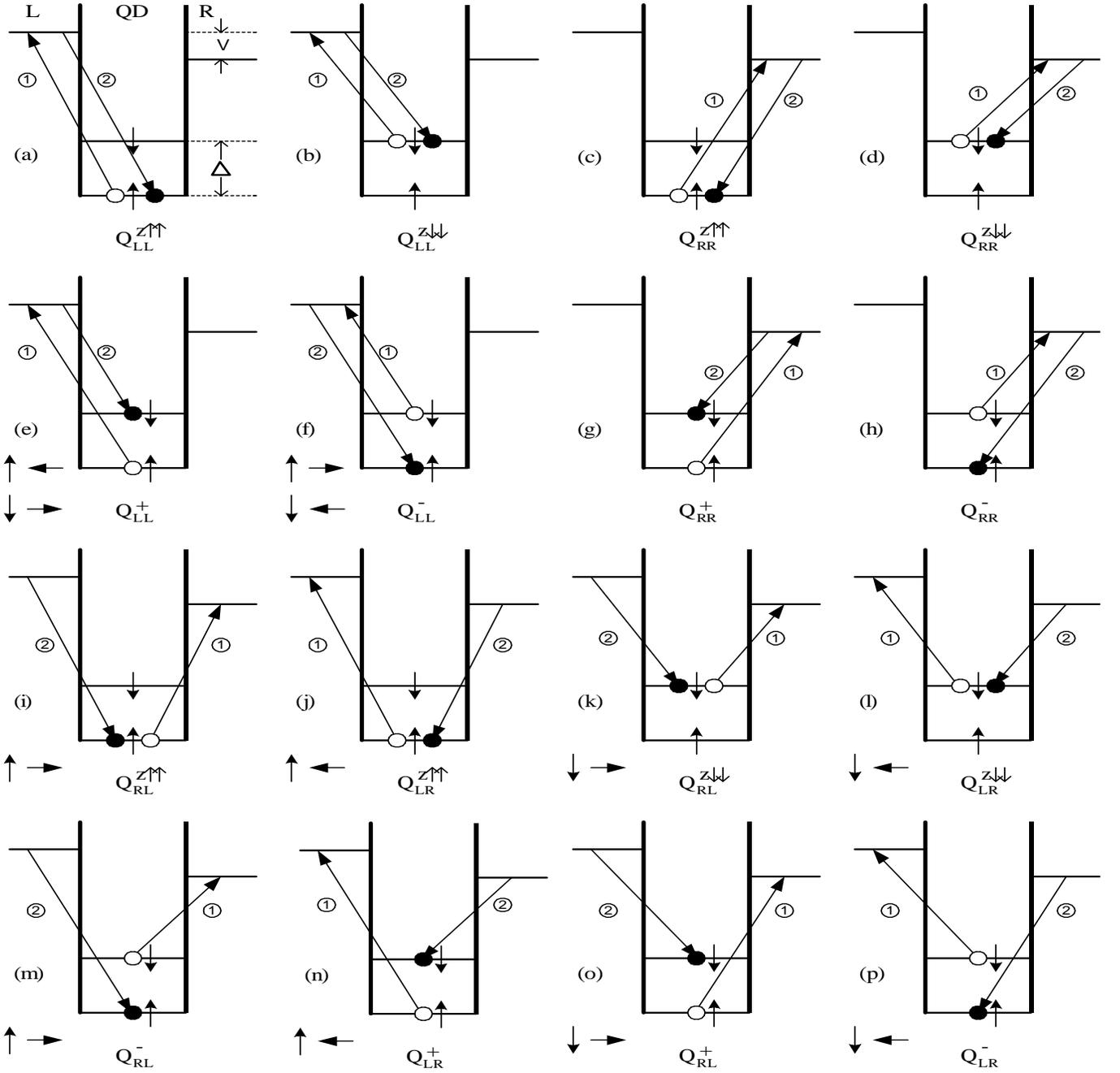}
\caption{Schematic description of all $16$ cotunneling processes (a)--(p) 
through a QD subject to Zeeman splitting energy $\Delta$ between the spin-up 
and -down electronic states, in the strong Coulomb blockade regime. A finite 
bias-voltage $V$ is applied between the two leads, L and R. An open circle 
inside the QD stands for the initial occupied electron state before tunneling 
events, while the solid circle denotes the final state occupied by an electron 
after the cotunneling processes. The single-particle tunneling event 
\textcircled{\footnotesize 1} takes place first and then is followed by the 
tunneling event \textcircled{\footnotesize 2}, which together comprise the 
entire cotunneling process. The reservoir variable below each of the figures 
denotes the corresponding physical process occurring in the reservoir. The 
arrow beneath the left lead in each of the figures denotes the flow direction 
of the spin-up and(or) spin-down electron with respect to the left lead.} 
\label{fig3}
\end{figure*}

Type-I cotunneling involves only one lead and equal spin orientations in two 
successive single-particle tunneling events. Figures 3(a)--(d) depict such 
trivial cotunneling events, in which an electron with spin-$\sigma$ exits the 
QD to the lead $\eta$ and subsequently an electron with the same spin in the 
same lead (probably not the same electron) transfers to the QD. Obviously, 
these events make no contribution to the current. In contrast, type-II 
cotunneling experiences a spin-flip process as shown in Fig.~3(e)--(h), in 
which the final spin state in the QD is opposite to its initial state. Of 
course, only events (e) and (f) relate to the currents in the left lead, and 
involve the terms $\mp Q_{LL}^- S^+$ and $\pm Q_{LL}^+ S^-$ in 
Eqs.~(\ref{iup}) and (\ref{idown}). However, the spin-up and spin-down 
electrons flow in opposite directions and contribute to the currents with 
equal magnitudes. As a result, no charge current occurs in this type 
cotunneling but spin current does emerge. Furthermore, because only one lead 
is involved, the contributions of events (e) and (f) are naturally independent 
of the bias-voltage and are only dependent on the magnetic field, involving 
the terms $\pm \pi \Delta J_{LL}^2 \rho_L^2 \mp 2\pi T J_{LL}^2 \rho_L^2 
\varphi(\Delta/T) S^z$ in spin-up and spin-down currents with $\pm\rightarrow 
+,-$, respectively. Figures 3(i)--(l) represent all type-III cotunneling 
processes. This kind of cotunneling describes an equivalent spin-conservative 
tunneling process, in which an electron is transferred from one lead to 
another lead via the QD without spin exchange. The corresponding terms in 
Eqs.~(\ref{iup}) and (\ref{idown}) are $\pm Q_{LR(RL)}^{z\uparrow\uparrow 
(\downarrow\downarrow)} S^z$. Moreover, we observe that (1) the spin-up events 
(i)\&(j) and the spin-down events (k)\&(l) yield currents having not only the 
same directions, but they also have equal magnitudes; and (2) these 
spin-up(down) events involve only the difference of chemical potentials, 
leading to a contribution proportional to the bias-voltage, ${3\over 2} \pi 
J_{LR}^2 \rho_L \rho_R V$. This observation reveals that type-III cotunneling 
excludes the possibility of spin current, but it does provide a linearly 
bias-voltage-dependent term in the charge current, Eq.~(\ref{ic}). Finally, 
type-IV cotunneling comprises the four electron-transferring tunneling events 
accompanied by a spin-flip process as exhibited in Figs.~3(m)--(p). They 
produce the terms $-Q_{RL}^{-}S^+$, $+Q_{LR}^+ S^-$ in Eq.~(\ref{iup}) and 
$Q_{LR}^- S^+$, $- Q_{RL}^+ S^-$ in Eq.~(\ref{idown}). Differing from type-III 
cotunneling, we find that in type-IV cotunneling the spin-up(down) events 
(m)\&(n) [or (o)\&(p)] involve both the voltage change and spin-flip, and the 
corresponding contributions to current are dependent on both $V$ and $\Delta$: 
$\pm [\Delta- 2 S^z T \varphi(\Delta\pm V/T)] \pi J_{LR}^2 \rho_L \rho_R$. 
Type-IV cotunneling produces both spin and charge currents. In sum, the 
mechanism for creating spin current stems solely from inelastic spin-flip 
cotunneling processes (type-II and IV), while the electron-transferring 
elastic and inelastic cotunneling processes (type-III and IV) are responsible 
for producing charge current.

Substituting the steady-state solution, Eq.~(\ref{magnetization}), into the 
charge current, Eq.~(\ref{ic}), and the spin current, Eq.~(\ref{is}), we 
readily find that: (1) both the charge current and the spin current are zero 
if $V=0$; (2) the resulting spin current is nonzero in nonequilibrium 
conditions, $V\neq 0$, if and only if two conditions are satisfied: 
$J_{LL}\neq J_{RR}$, i.e. the asymmetrical Kondo coupling case, and there is a 
nonvanishing magnetic field, $\Delta \neq 0$; (3) the spin current is an even 
function of the applied bias-voltage, $V$, indicating that the sign of spin 
current is not related to the direction of the bias-voltage; whereas charge 
current, Eq.~(\ref{ic}), is an odd function of the bias-voltage and will 
change its sign when bias-voltage is applied in the opposite direction; (4) 
the magnetic-field-related spin current changes its sign when the direction of 
the applied magnetic field is reversed (odd function), while the charge 
current is an even function of $\Delta$, because it measures the results of 
the total charge flow irrespective of the spin orientation. The sign property 
of the spin current was also pointed out in previous study.\cite{Schiller}

As an illustration, we plot the bias-voltage dependent differential 
conductance, $dI^c/dV$, in Fig.~4. The differential conductance shows a 
characteristic jump at $V=\pm \Delta$, which is the signature of the Kondo 
effect in the presence of an external magnetic field. Mathematically, this 
feature comes from the hyperbolic cotangent function in the current formula, 
Eq.~(\ref{ic}). From a physical point of view, this splitting can be 
qualitatively understood from the following consideration: a small 
bias-voltage, $|V|< \Delta$, can not provide enough energy to spur the 
spin-flip cotunneling process that is an energy-consuming event in the case of 
nonzero magnetic field; However, when $|V| \geq \Delta$, the spin-flip 
cotunneling process is energetically activated, thus an additional channel is 
opened for electron transport. Moreover, the effect of temperature is to smear 
and reduce the peak.

\begin{figure}[htb]
\includegraphics [width=7cm,height=5cm,angle=0,clip=on]{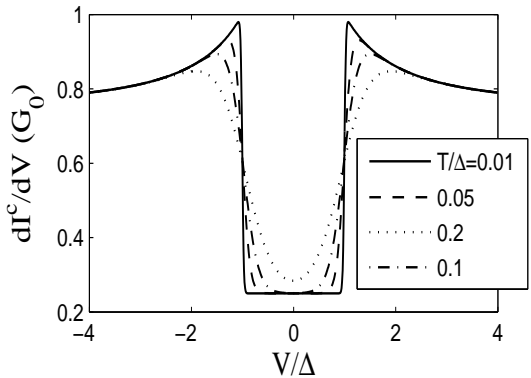}
\caption{The calculated differential conductance $dI^c/dV$ vs. bias-voltage 
$V/\Delta$ for several temperatures at nonzero magnetic field in units of 
$G_0=4\pi J_{LR}^2 \rho_L \rho_R$ (linear conductance at zero magnetic field). 
Other parameters are the same as in Fig.~1.} \label{fig4}
\end{figure}

\begin{figure}[htb]
\includegraphics [width=8.5cm,height=8cm,angle=0,clip=on]{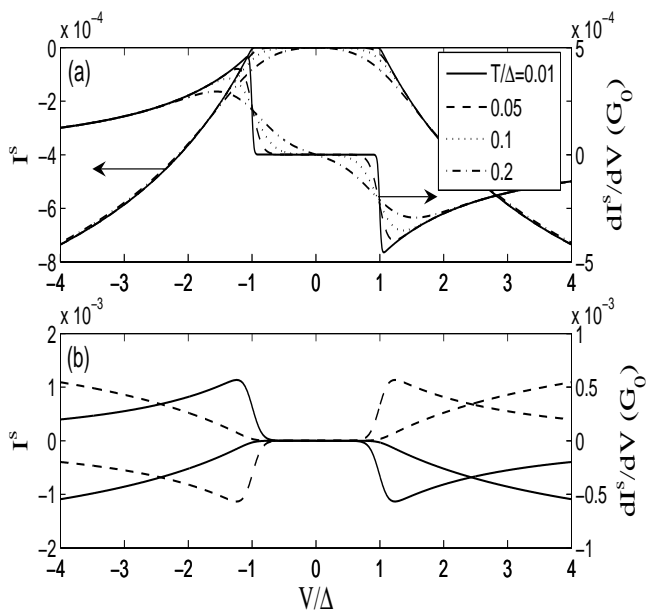}
\caption{The calculated spin current, $I^s$, and its differential conductance, 
$dI^s/dV$, as functions of bias-voltage $V/\Delta$, at nonzero magnetic field. 
(a) exhibits results for several temperatures and $J_{RR}/J_{LL}=4.0$, 
$J_{LL}=0.02$. (b) plots the results for $J_{RR}/J_{LL}=5.0$ ($J_{LL}=0.02$) 
as solid lines, and for $J_{LL}/J_{RR}=5.0$ ($J_{RR}=0.02$) as dashed lines.} 
\label{fig5}
\end{figure}

We also exhibit the resulting spin current, $I^s$, and its differential 
conductance, defined as $dI^s/dV$, as functions of bias-voltage $V/\Delta$ in 
Fig.~5. We observe that at low temperatures, $T/\Delta=0.01$ and $0.05$ in 
Fig.~5(a), the calculated spin currents are nearly zero in the small 
bias-voltage region, $|V|< \Delta$, notwithstanding $J_{LL}/J_{RR}=4.0$ and 
$\Delta \neq 0$. Analogous to the peak-splitting of the differential 
conductance shown in Fig.~4, the low-temperature vanishing of spin current is 
also due to the fact that spin-flip scattering is energetically inaccessible 
in the case of small bias-voltage. This vanishing produces a ``window" of zero 
differential conductance for spin current. Nevertheless, at higher 
temperatures, thermal fluctuation provides an additional possibility to flip 
spin in the tunneling processes, leading to a slow increase of the spin 
current and the gradual disappearance of the zero ``window" in $dI^s/dV$. In 
Fig.~5(b), we show that the sign of the spin current is also determined by the 
relative magnitudes of $J_{LL}$ and $J_{RR}$.

\section{Conclusions}\label{s:sum}

In this paper, we have systematically examined nonequilibrium inelastic 
cotunneling through a single spin (QD) subject to a finite magnetic field in 
the strong Coulomb blockade regime, in the weak tunnel coupling limit. For 
this purpose, we introduced the Kondo Hamiltonian to model cotunneling in the 
QD and employed a generic Heisenberg-Langevin equation approach to establish a 
set of quantum Bloch-type dynamical equations describing inelastic cotunneling 
phenomenology.

In our formulation, the operators of the localized spin and the reservoirs 
were first determined formally by integration of their Heisenberg equations of 
motion, exactly to all orders in the tunnel coupling constants. Next, under 
the assumption that the time scale of the decay processes is much slower than 
that of free evolution, we expressed the time-dependent operators involved in 
the integrands of these equations of motion approximately in terms of their 
free evolution. Thirdly, these equations of motion were expanded in powers of 
the tunnel-coupling constants to second order; this approximation is 
physically valid in the weak tunnel-coupling limit. On the basis of these 
consideration, jointly with normal ordering, we developed the Bloch-type 
equations expressed explicitly and compactly in terms of the response and the 
correlation functions of the free reservoir variables, which facilitated our 
theoretical examination of relaxation and decoherence in the localized spin 
induced by the ``environment".

In the problem at hand, dissipation of the QD spin stems from tunnel-coupling 
of the QD to two leads by cotunneling mechanisms. Based on our derived Bloch 
equations, we obtained explicit analytical expressions for the corresponding 
relaxation and decoherence rates at arbitrary bias-voltage and temperature. We 
found that relaxation results exclusively from spin-flip cotunneling processes 
alone, whereas both spin-flip and non-spin-flip cotunneling events contribute 
to decoherence. In this analysis, we carried out systematic examinations of 
the relaxation rate and the decoherence rate as functions of bias-voltage, 
external magnetic field, and temperature. Our formulation also facilitated the 
derivation of an analytic expression for the nonequilibrium magnetization that 
is found to match that of earlier theories.

Employing this approach, we also derived closed-form expressions for the 
spin-resolved currents, which facilitated our calculation of both the charge 
current and the spin current. Furthermore, we classified and examined all 
possible cotunneling processes occurring in the strong Coulomb interaction QD 
($16$ events), and categorized them in  four distinct types. In this, we found 
that: (1) type-I cotunneling make no contribution to current; (2) spin-flip 
processes, types-II and IV cotunneling, drive the spin current; (3) the 
electron-transferring processes, types-III (non-spin-flip) and IV (spin-flip) 
cotunneling, produce charge current; and (4) we also determined formulae for 
their respective contributions to the charge and spin currents. Our numerical 
calculations exhibit splitting of the zero-bias-voltage peak in the 
differential conductance for charge current in a finite magnetic field, which 
is a typical signature of the Kondo effect, and a wide ``window" of zero 
differential conductance for spin current about zero-bias-voltage.
With insight gained from our specific analyses, we can ascribe these 
low-temperature transport characteristics to the fact that {\em inelastic} 
spin-flip cotunneling is energetically active only for sufficiently strong 
applied bias-voltage, $V\geq\Delta$. We have also shown that spin current, 
unlike charge current, is an even function of the applied bias-voltage, and 
its direction depends on the orientation of the ambient magnetic field and 
asymmetry of the Kondo coupling constants to the left and the right leads.

\begin{acknowledgments}

This work was supported by Department of Defense through the DURINT Program 
administered by the US Army Research Office, DAAD Grant No.19-01-1-0592.

\end{acknowledgments}

\appendix

\section{Derivation of Eqs.~(\ref{Q:qz}), (\ref{eom:sz1}), and 
(\ref{eom:s+1})}\label{app:a}

In this appendix, we first prove Eq.~(\ref{Q}). Consider $Q_{\eta\eta'}^{z}$ 
for example. Substituting Eq.~(\ref{coi}) into the definition of 
Eq.~(\ref{Qz}), we have
\bn
Q_{\eta\eta'}^{z}(t)&=& J_{\eta\eta'} \sum_{{\bf k}, {\bf k}'} \bigl [ c_{\eta 
{\bf k} \uparrow}^{o\dagger}(t)+ c_{\eta {\bf k} \uparrow}^{i\dagger}(t) \bigr 
] \cr
&& \hspace{-1cm} \times \bigl [ c_{\eta' {\bf k}' \uparrow}^{o}(t) + c_{\eta' 
{\bf k}' \uparrow}^{i}(t) \bigr ] - J_{\eta\eta'} \sum_{{\bf k}, {\bf k}'} 
(\uparrow\rightarrow\downarrow) \cr
&&\hspace{-1cm}  =Q_{\eta\eta'}^{zo}(t) -i J_{\eta\eta'} \sum_{{\bf k}, {\bf 
k}'} \int_0^t dt' \cr
&& \times \left \{ c_{\eta {\bf k} \uparrow}^{o\dagger}(t) : [c_{\eta' {\bf 
k}' \uparrow}^o(t), H_{\rm I}^o(t')]_- : \right. \cr
&& \hspace{-1cm} \left. - : [c_{\eta {\bf k} \uparrow}^{o\dagger}(t), H_{\rm 
I}^o(t')]_- : c_{\eta' {\bf k}' \uparrow}^{o}(t)\right \} + iJ_{\eta\eta'} 
\sum_{{\bf k},{\bf k}'} (\uparrow\rightarrow\downarrow) \cr
&& \hspace{-1cm}  =Q_{\eta\eta'}^{zo}(t) -i J_{\eta\eta'} \sum_{{\bf k},{\bf 
k}'} \theta(\tau) \int_{0}^\infty dt' \cr
&& \hspace{-1cm}\times : [c_{\eta {\bf k} \uparrow}^{o\dagger}(t) c_{\eta' 
{\bf k}' \uparrow}^o(t), H_{\rm I}^o(t')]_- : + iJ_{\eta\eta'} \sum_{{\bf 
k},{\bf k}'} (\uparrow\rightarrow\downarrow) \cr
&&\hspace{-1cm}  =Q_{\eta\eta'}^{zo}(t) - i\theta(\tau) \int_{-\infty}^t d\tau 
: [Q_{\eta\eta'}^{zo}(t), H_{\rm I}^o(t')]_-:\,.\cr
&& \label{appa:qz}
\en
In the second stage of Eq.~(\ref{appa:qz}), we neglect terms of the form, 
$c_{\eta {\bf k} \sigma}^{i\dagger} c_{\eta' {\bf k}' \sigma'}^{i}$, since 
they are of second-order in the coupling constant, $O(J^2)$, yielding a 
third-order contribution to $Q_{\eta\eta'}^z$ with respect to $J$.

To derive Eqs.~(\ref{eom:sz1}) and (\ref{eom:s+1}), we consider the normally 
ordered product of the reservoir and spin operators, $:~Q^{a}S^{b}:$,
\begin{align}
& :Q^a(t)S^b(t):= :[Q^{a}_o(t) + Q^{a}_i(t)] [S_o^b(t) + S_{i}^b(t)]: \cr
&=\, :Q^{a}_o(t) S_{o}^b(t): -i\theta(\tau) \int_{-\infty}^t d\tau \left 
\{:Q^{a}_o(t) \right. \cr
&\, \times : [S_{o}^b(t), H_{\rm I}^o(t')]_-: \left. + : [Q^{a}_o(t), H_{\rm 
I}^o(t')]_-: S_o^b(t): \right \}. \cr
& \label{qasb}
\end{align}
Once again, we neglect the term $Q^a_i S_i^b$ as it is proportional to 
$O(J^3)$. The first term in Eq.~(\ref{qasb}) involves only the free reservoir 
variables and the decoupled single spin. The other interaction terms (we 
designate the operator expressions in the integrand as ${\cal I}$) arise from 
tunneling reaction, upon which we focus in the following derivation. Using the 
compact definition of the interaction Hamiltonian, Eq.~(\ref{Hi}), $H_{\rm 
I}=\sum_{c\in \{ z,+,-\}} Q^{c} F_{Q^c}$, we have
\bn
{\cal I}&=& \sum_{c} \bigl \{: Q_o^a(t) Q_o^c(t') [S_o^b(t), F_{Q^c}^o(t')]_-  
\cr
&& + [Q_o^a(t), Q_o^c(t')]_- F_{Q^c}^o(t') S_o^b(t) : \bigr \} \cr
&=& \sum_{c} \left \{ :{1\over 2} [Q_o^a(t), Q_o^c(t')]_+ [S_o^b(t), 
F_{Q^c}^o(t')]_-  \right. \cr
&& + {1\over 2} [Q_o^a(t), Q_o^c(t')]_- [S_o^b(t), F_{Q^c}^o(t')]_-  \cr
&& \left. + [Q_o^a(t), Q_o^c(t')]_- F_{Q^c}^o(t') S_o^b(t) : \right \} \cr
&=& {1\over 2} \sum_{c} \left \{ : [Q_o^a(t), Q_o^c(t')]_+ [S_o^b(t), 
F_{Q^c}^o(t')]_- \right. \cr
&& \left. + [Q_o^a(t), Q_o^c(t')]_- [S_o^b(t), F_{Q^c}^o(t')]_+ : \right \}.
\en
Therefore, the full normal-ordered operator product ($:~\!\! Q^a S^b:$) is 
written as the sum of a zero-order term and a term of second-order in the 
coupling constant $J$, having the compact form:
\bn
&& \hspace{-0.5cm} :Q^a(t)S^b(t):= :Q^{a}_o(t) S_{o}^b(t): -i \int_{-\infty}^t 
d\tau \sum_{c} \bigl \{: \hat C^{ac}(t,t') \cr
&& \hspace{-0.5cm} \times [S_o^b(t), F_{Q^c}^o(t')]_- + \hat R^{ac}(t,t') 
[S_o^b(t), F_{Q^c}^o(t')]_+ : \bigr \},
\en
with the definitions
\bn
\hat R^{ac}(t,t') &=& {1\over 2}\theta(\tau) [Q_o^a(t), Q_o^c(t')]_-, \\
\hat C^{ac}(t,t') &=& {1\over 2}\theta(\tau) [Q_o^a(t), Q_o^c(t')]_+ .
\en
The reservoir equilibrium ensemble averages of $\hat R^{ac}(t,t')$ and $\hat 
C^{ac}(t,t')$ are just the response function $R^{ac}(t,t')$ and the 
correlation function $C^{ac}(t,t')$ defined in Eqs.~(\ref{responsef}) and 
(\ref{correlationf}), respectively.

In the spin operator equations of motion, these zero-order terms contribute 
quantum fluctuations associated with the reservoir fields, as well as quantum 
effects pertaining to the intrinsic character of the reservoirs (for example, 
superconducting or ferromagnetic leads). In any event, we take the ensemble 
average of each equation of motion separately in regard to the electron 
ensembles of the reservoirs and in regard to the quantum spin states. Thus, 
the normally ordered operator products factorize in the averaging procedure. 
Considering that we take no account of quantum fluctuations in the present 
paper and only normal leads are connected to the single spin, the zero-order 
terms make no contribution to the quantum Bloch equations. Moreover, only 
$\langle Q_o^{\pm}(t) Q_o^{\mp}(t') \rangle_e$ and $\langle Q_o^{z}(t) 
Q_o^{z}(t') \rangle_e$ are nonzero for normal leads (see Appendix 
\ref{app:b}). Combining all the above results, we obtain Eqs.~(\ref{eom:sz1}) 
and (\ref{eom:s+1}).

\section{Response and Correlation functions of the reservoirs}\label{app:b}

To obtain explicit expressions for the Bloch equations and the current, we 
need to determine the various correlation and response functions $C(R)_{\eta_1 
\eta_2,\eta_3 \eta_4}^{ab}(t,t')$ of the free reservoir variables, which are 
defined as
\bq
C(R)_{\eta_1 \eta_2,\eta_3 \eta_4}^{ab}(t,t')={1\over 2}\theta(\tau) \langle 
[Q_{\eta_1 \eta_2}^{ao}(t), Q_{\eta_3 \eta_4}^{bo}(t')]_{\pm} \rangle.
\eq

In the following, we drop all super(sub)scripts, ``$o$", bearing in mind that 
all operators are free reservoir operators. In our calculations, we assume 
that: (i) the leads have a flat density of states $\rho_{\eta}$ for both spin 
orientations, so we can make the replacement
\bq
\sum_{{\bf k}} (\cdots) \longrightarrow \rho_{\eta} \int d\epsilon (\cdots);
\eq
(ii) the normal leads are in the respective bias-voltage driven local 
equilibrium states described by
\bq
f_{\eta}(\epsilon)=\left [1+ e^{(\epsilon-\mu_{\eta})/T} \right ]^{-1},
\eq
with temperature $T$ and chemical potential $\mu_{\eta}$; and (iii) the time 
evolution of free reservoir operators is governed by Eq.~(\ref{free:c}). 
According to Wick's theorem and properties (i) and (ii), it is easy to see 
that only the functions $C(R)_{\eta\eta',\eta'\eta}^{zz}$, 
$C(R)_{\eta\eta',\eta'\eta}^{+-}$, and $C(R)_{\eta\eta',\eta'\eta}^{-+}$ are 
nonzero. We calculate them individually.

1. $C(R)^{zz}(t,t')=\sum_{\eta, \eta'}C(R)_{\eta\eta', \eta'\eta}^{zz}(t,t')$:
\begin{align}
& C(R)_{LR,RL}^{zz\uparrow\uparrow}(t,t')={1\over 2}\theta(\tau) \langle 
[Q_{LR}^{z\uparrow\uparrow}(t), Q_{RL}^{z\uparrow\uparrow}(t')]_\pm \rangle_e 
\cr
&\,= {1\over 2} \theta(\tau) J_{LR}^2 \sum_{{\bf k},{\bf k}',{\bf q},{\bf q}'} 
\langle [c_{L{\bf k} \uparrow}^\dagger (t) c_{R {\bf k}' \uparrow}^\pdag (t), 
c_{R {\bf q} \uparrow}^\dagger(t') c_{L {\bf q}' \uparrow}^\pdag (t')]_\pm 
\rangle_e \cr
&\,= {1\over 2} \theta(\tau) J_{LR}^2 \sum_{{\bf k},{\bf k}',{\bf q},{\bf q}'} 
e^{i (\epsilon_{L{\bf q}'}- \epsilon_{R {\bf q}})\tau } \cr
&\, \times \left [ \langle c_{L {\bf k} \uparrow}^\dagger (t) c_{L {\bf q}' 
\uparrow}^\pdag (t) \rangle_e \langle c_{R {\bf k}' \uparrow}^\pdag (t) c_{R 
{\bf q} \uparrow}^\dagger (t) \rangle_e  \right. \cr
&\, \left. \pm \langle c_{R {\bf q} \uparrow}^\dagger (t) c_{R {\bf k}' 
\uparrow}^\pdag (t) \rangle_e \langle c_{L {\bf q}' \uparrow}^\pdag (t) c_{L 
{\bf k} \uparrow}^\dagger (t) \rangle_e \right ]\cr
&\,= {1\over 2} \theta(\tau) J_{LR}^2 \rho_L \rho_R \int d\epsilon d\epsilon' 
e^{i(\epsilon-\epsilon')\tau} \cr
&\, \times \left \{ f_{L}(\epsilon) \left [ 1-f_{R}(\epsilon') \right ] \pm 
f_{R}(\epsilon') \left [ 1-f_{L}(\epsilon) \right ] \right \}. \label{crzz1}
\end{align}
Exchanging the roles of $R$ and $L$, $L\leftrightarrow R$, 
$C(R)_{RL,LR}^{zz\uparrow\uparrow}(t,t')$ yields
\bn
C(R)_{RL,LR}^{zz\uparrow\uparrow}(t,t')&=&{1\over 2} \theta(\tau) J_{LR}^2 
\rho_L \rho_R \int d\epsilon d\epsilon' e^{-i(\epsilon-\epsilon')\tau} \cr
&&\hspace{-3cm} \times \left \{ f_{R}(\epsilon') \left [ 1-f_{L}(\epsilon) 
\right ] \pm f_{L}(\epsilon) \left [ 1-f_{R}(\epsilon') \right ] \right \}. 
\label{crzz2}
\en
As we take the leads to be normal metals/semiconductors, we have
\begin{eqnarray*}
C(R)_{LR,RL}^{zz\downarrow\downarrow}(t,t')&=&{1\over 2}\theta(\tau) \langle 
[Q_{LR}^{z\downarrow\downarrow}(t), Q_{RL}^{z\downarrow\downarrow}(t')]_\pm 
\rangle_e \cr
&=& C(R)_{LR,RL}^{zz\uparrow\uparrow}(t,t'), \cr
C(R)_{RL,LR}^{zz\downarrow\downarrow}(t,t')&=& 
C(R)_{RL,LR}^{zz\uparrow\uparrow}(t,t'), \cr
C(R)_{LR,RL}^{zz\uparrow\downarrow}(t,t')&=&{1\over 2}\theta(\tau) \langle 
[Q_{LR}^{z\uparrow\uparrow}(t), Q_{RL}^{z\downarrow\downarrow}(t')]_\pm 
\rangle_e=0, \cr
C(R)_{LR,RL}^{zz\uparrow\downarrow}(t,t')&=&0.
\end{eqnarray*}
Furthermore,
\begin{align}
& C(R)_{LL,LL}^{zz}(t,t')={1\over 2}\theta(\tau) \langle [Q_{LL}^{z}(t), 
Q_{LL}^{z}(t')]_\pm \rangle_e \cr
&\,= \theta(\tau) J_{LL}^2 \rho_L^2 \int d\epsilon d\epsilon' 
e^{i(\epsilon-\epsilon')\tau} \cr
&\, \times \left \{ f_{L}(\epsilon) \left [ 1-f_{L}(\epsilon') \right ] \pm 
f_{L}(\epsilon') \left [ 1-f_{L}(\epsilon) \right ] \right \},
\end{align}
\begin{align}
& C(R)_{RR,RR}^{zz}(t,t')={1\over 2}\theta(\tau) \langle [Q_{RR}^{z}(t), 
Q_{RR}^{z}(t')]_\pm \rangle_e \cr
&\,= \theta(\tau) J_{RR}^2 \rho_R^2 \int d\epsilon d\epsilon' 
e^{i(\epsilon-\epsilon')\tau} \cr
&\, \times \left \{ f_{R}(\epsilon) \left [ 1-f_{R}(\epsilon') \right ] \pm 
f_{R}(\epsilon') \left [ 1-f_{R}(\epsilon) \right ] \right \}.
\end{align}
Finally, $C(R)^{zz}(t,t')$ are functions only of the time difference $\tau$ 
and take the form:
\bn
&& C(R)^{zz}(\tau) = \theta(\tau) \sum_{\eta} J_{\eta \eta}^2 \rho_\eta^2 \int 
d\epsilon d\epsilon' e^{i(\epsilon-\epsilon') \tau} \cr
&& \times \left \{ f_{\eta}(\epsilon) \left [ 1-f_{\eta}(\epsilon') \right ] 
\pm f_{\eta}(\epsilon') \left [ 1-f_{\eta}(\epsilon) \right ] \right \} \cr
&& + \theta(\tau) J_{LR}^2 \rho_L \rho_R \int d\epsilon d\epsilon' \left [ 
e^{i(\epsilon-\epsilon')\tau} \pm  e^{-i(\epsilon-\epsilon')\tau} \right ] \cr
&& \times \left \{ f_{L}(\epsilon) \left [ 1-f_{R}(\epsilon') \right ] \pm 
f_{R}(\epsilon') \left [ 1-f_{L}(\epsilon) \right ] \right \}.
\en

2. $C(R)^{+-/-+}(t,t')=\sum_{\eta,\eta'} C(R)_{\eta\eta', 
\eta'\eta}^{+-/-+}(t,t')$:
\begin{align}
& C(R)_{LR,RL}^{+-/-+}(\tau)={1\over 2}\theta(\tau) \langle [Q_{LR}^{+/-}(t), 
Q_{RL}^{-/+}(t')]_\pm \rangle_e \cr
&\,={1\over 2} \theta(\tau) J_{LR}^2 \rho_L \rho_R \int d\epsilon d\epsilon' 
e^{i(\epsilon-\epsilon')\tau} \cr
&\,\,\,\,\,\, \times \left \{ f_{L}(\epsilon) \left [ 1-f_{R}(\epsilon') 
\right ] \pm f_{R}(\epsilon') \left [ 1-f_{L}(\epsilon) \right ] \right \}, 
\label{cr+-1}
\end{align}
\begin{align}
& C(R)_{RL,LR}^{+-/-+}(\tau)={1\over 2}\theta(\tau) \langle [Q_{RL}^{+/-}(t), 
Q_{LR}^{-/+}(t')]_\pm \rangle_e \cr
&\,={1\over 2} \theta(\tau) J_{LR}^2 \rho_L \rho_R \int d\epsilon d\epsilon' 
e^{i(\epsilon-\epsilon')\tau} \cr
&\,\,\,\,\,\, \times \left \{ f_{R}(\epsilon) \left [ 1-f_{L}(\epsilon') 
\right ] \pm f_{L}(\epsilon') \left [ 1-f_{R}(\epsilon) \right ] \right \}, 
\label{cr+-2}
\end{align}
\bn
&& C(R)_{LL,LL \atop RR,RR}^{+-/-+}(\tau)= {1\over 2}\theta(\tau) \langle 
[Q_{LL \atop RR}^{+/-}(t), Q_{LL \atop RR}^{-/+}(t')]_\pm \rangle_e \cr
&&= {1\over 2} \theta(\tau) J_{LL/RR}^2 \rho_{L/R}^2 \int d\epsilon d\epsilon' 
e^{i(\epsilon-\epsilon')\tau} \cr
&&\times \left \{ f_{L/R}(\epsilon) \left [ 1-f_{L/R}(\epsilon') \right ] \pm 
f_{L/R}(\epsilon') \left [ 1-f_{L/R}(\epsilon) \right ] \right \}. \cr
&& \label{cr+-3}
\en
Moreover, we can easily obtain
\bq
C(R)^{+-}(\tau)=C(R)^{-+}(\tau)={1\over 2} C(R)^{zz}(\tau).
\eq

Therefore, in $\omega$-Fourier space, the spectral function $C(\omega)$ 
defined in Eq.~(\ref{ctau}) is given by
\bn
C(\omega)&=& {1\over 2}\int_{-\infty}^\infty d\tau e^{i\omega \tau} 
C^{zz}(\tau) \cr
&=& {1\over 2} \pi \sum_{\eta} J_{\eta\eta}^2 \rho_\eta^2 \int d\epsilon \bigl 
\{ f_\eta (\epsilon) \left [ 1- f_\eta (\epsilon+ \omega) \right ] \cr
&& + f_\eta (\epsilon+ \omega) \left [ 1- f_\eta (\epsilon) \right ] \bigr \} 
\cr
&& + {1\over 2} \pi J_{LR}^2 \rho_L \rho_R \int d\epsilon \sum_{\eta} \bigl \{ 
f_{\eta}(\epsilon) \bigl [ 1- f_{\bar \eta}(\epsilon+ \omega) \bigr ] \cr
&& + f_{\eta}(\epsilon) \bigl [ 1- f_{\bar \eta}(\epsilon- \omega) \bigr ] 
\bigr \}.
\en
Also, the imaginary part of the retarded susceptibility $R(\omega)$ defined in 
Eq.~(\ref{rtau}) is
\bn
R(\omega)&=& {1\over 2}\int_{-\infty}^\infty d\tau e^{i\omega \tau} 
R^{zz}(\tau) \cr
&=& {1\over 2} \pi \sum_{\eta} J_{\eta\eta}^2 \rho_\eta^2 \int d\epsilon \bigl 
[ f_\eta (\epsilon) - f_\eta (\epsilon + \omega) \bigr ] \cr
&& + {1\over 2} \pi J_{LR}^2 \rho_{L} \rho_{R} \int d\epsilon \bigl [ 
f_R(\epsilon- \omega) - f_R(\epsilon + \omega) \bigr ]. \cr
&&
\en
It is readily seen that
\bq
C(-\omega)=C(\omega),\,\,\,\,\,R(-\omega)=-R(\omega).
\eq
Using the formulae
\bq
\int d\epsilon [f_{\eta}(\epsilon + \omega) - f_{\eta'}(\epsilon)]= 
-(\mu_{\eta'} - \mu_\eta + \omega), \label{fermi:integral1}
\eq
and
\bq
\int d\epsilon f_{\eta}(\epsilon+\omega) [1- f_{\eta'}(\epsilon)]= 
\frac{\omega - \mu_{\eta}+ \mu_{\eta'}}{e^{(\omega-\mu_{\eta}+ \mu_{\eta'})/T} 
- 1}, \label{fermi:integral2}
\eq
we can perform the $\epsilon$-integrals in the large bandwidth limit, with the 
results
\bn
R(\omega)&=& \pi \left ( \frac{J_{LL}^2 \rho_L^2 + J_{RR}^2 \rho_R^2}{2} + 
J_{LR}^2 \rho_{L} \rho_{R} \right ) \omega, \label{response} \\
C(\omega)&=& \frac{\pi}{2} \left ( J_{LL}^2 \rho_L^2+ J_{RR}^2 \rho_R^2 \right 
) T \varphi \left ( \frac{\omega}{T} \right ) \cr
&& + \frac{\pi}{2} J_{LR}^2 \rho_{L}\rho_{R} T \left [ \varphi \left ( 
\frac{\omega + V}{T}\right ) + \varphi \left ( \frac{\omega- V}{T}\right ) 
\right ],\cr
&& \label{correlation}
\en
where we have defined
\bq
\varphi(x)\equiv x \coth \left ( \frac{x}{2}\right ). \label{phi}
\eq
It should be noted that when the leads are in thermodynamic equilibrium, 
$\mu_{L}=\mu_{R}$, the spectral function $C(\omega)$ and the function 
$R(\omega)$ obey the fluctuation-dissipation theorem $C(\omega)=R(\omega) 
\coth(\omega/2T)$.

\end{document}